\documentclass[aps,superscriptaddress,twocolumn,prl]{revtex4-1}

\usepackage{graphicx,epsfig}
\usepackage{color}
\usepackage[usenames,dvipsnames]{xcolor}
\usepackage{amsmath,bbm,amssymb, amsthm}
\usepackage{dsfont} 
\usepackage{stmaryrd}
\usepackage{numprint}

\def\bb{\begin{eqnarray}}
\def\ee{\end{eqnarray}}
\newcommand{\ket}[1]{| #1 \rangle}
\newcommand{\bra}[1]{\langle #1 |}

\newcommand{\moy}[1]{\left\langle #1 \right\rangle} 
\newcommand{\idop}{\mathds{1}}

\renewcommand{\r}{\textbf{r}}
\newcommand{\Q}{{\cal Q}}
\newcommand{\x}{\textbf{x}}

\begin{document}

\title{Fluctuation Theorems for Continuous Quantum Measurement and Absolute Irreversibility}

\author{Sreenath K. Manikandan}
\affiliation{Department of Physics and Astronomy, University of Rochester, Rochester, NY 14627, USA}
\affiliation{Center for Coherence and Quantum Optics, University of Rochester, Rochester, NY 14627, USA}
\author{Cyril Elouard}
\affiliation{Department of Physics and Astronomy, University of Rochester, Rochester, NY 14627, USA}
\affiliation{Center for Coherence and Quantum Optics, University of Rochester, Rochester, NY 14627, USA}
\author{Andrew N. Jordan}
\affiliation{Department of Physics and Astronomy, University of Rochester, Rochester, NY 14627, USA}
\affiliation{Center for Coherence and Quantum Optics, University of Rochester, Rochester, NY 14627, USA}
\affiliation{Center for Quantum Studies, Chapman University, Orange, CA, USA, 92866}

\date{\today}

\begin{abstract}
Fluctuation theorems are relations constraining the out-of-equilibrium fluctuations of thermodynamic quantities like the entropy production that were initially introduced for classical or quantum systems in contact with a thermal bath. Here we show, in the absence of thermal bath, the dynamics of continuously measured quantum systems can also be described by a fluctuation theorem, expressed in terms of a recently introduced arrow of time measure. This theorem captures the emergence of irreversible behavior from microscopic reversibility in continuous quantum measurements. From this relation, we demonstrate that measurement-induced wave-function collapse exhibits absolute irreversibility, such that Jarzynski and Crooks-like equalities are violated. We apply our results to different continuous measurement schemes on a qubit: dispersive measurement, homodyne and heterodyne detection of a qubit's fluorescence.
\end{abstract}
\maketitle
The emergence of macroscopic irreversibility from microscopic time-reversal invariant physical laws has been a long-standing issue, well described by the formalism of statistical thermodynamics \cite{Sekimoto10,Seifert08}. In this framework, the small system under study follows stochastic trajectories in its phase-space, where the randomness models the uncontrolled forces exerted on the system by its thermal environment. Although these trajectories are microscopically reversible, one direction of time is more probable than the other and a arrow of time emerges for the ensemble of trajectories. In this framework, the thermodynamic variables like the work, the heat and the entropy produced during a process appear as random variables, defined for a single realization (i.e. a single trajectory), whose averages comply with the first and second law of thermodynamics. Furthermore, the fluctuations of these quantities are constrained beyond the second law, as captured by the so-called Fluctuation Theorems (FT)~\cite{Jarzynski97,Crooks99,Seifert05}, which can be written under the form $\moy{e^{-\sigma(\Gamma)}}= 1$, where $\sigma(\Gamma)$ is the entropy production along a single trajectory $\Gamma$. We denote $\moy{\cdot}$, the ensemble average over the realizations of the studied process (or equivalently, over the possible trajectories). The entropy production $\sigma(\Gamma)$ fulfilling the FT is equal to the ratio of the probability of the (forward in time) trajectory $\Gamma$ and the probability of the time-reversed (or backward in time) trajectory corresponding to $\Gamma$. During the last decades, these results have been investigated in the quantum regime where the system and the thermal bath can be quantum systems, allowing the proof of quantum extensions of the FTs~\cite{Kurchan01,Mukamel03,Campisi09,Esposito09,Campisi11,Chetrite12,Horowitz12,
Horowitz13,campisi2011influence,campisi2010fluctuation,miller2017time,Hekking13,Manzano15}.
Experiments have demonstrated the validity of these FTs in both classical and quantum regimes~\cite{Toyabe10,Saira12,Berut13,Batalhao14,Naghiloo18,Naghiloo17}. 

However, it was shown that the form of the FTs must be modified for special processes ~\cite{teifel2010limitations,sagawa2010generalized,Murashita14,morikuni2011quantum,Funo15,gross2005flaw,jarzynski2005reply,gross2005reply,quan2012validity,crooks2007work}, which are such that some theoretically allowed backward trajectories do not have a forward-in-time counterpart. A canonical example is the free expansion of a single particle gas initially contained in the left half of a box by a wall. The wall is removed at time $t=0$, letting the gas expand and reach thermal equilibrium in the whole box. The reverse process consists in starting with the gas particle equilibrated in the whole box and reinserting the wall in the middle. Half of the time, after putting back the wall, the gas particle will be found in the right half of the box. However, this configuration is forbidden in the initial state of the gas, and then only the realizations for which the particle is found in the left-hand side after reinserting the wall can be associated to a realization of the direct process~\cite{teifel2010limitations,Murashita14,sung2008application,crooks2007work}. For the general class of processes in which this phenomenon occurs, qualified as absolutely irreversible \cite{Murashita14,Funo15}, the FTs takes the form $\moy{e^{-\sigma(\Gamma)}}= 1-\lambda$, where $\lambda \in [0,1]$ is the accumulated probability of the backward trajectories with no forward counterparts. Absolutely irreversible processes exhibit a strictly positive average entropy production, bounded below by $-\log(1-\lambda)>0$. Reversibility, i.e. a zero average entropy production, is impossible for such processes, no matter the speed at which one implements the transformation under study.

Recently,  stochastic thermodynamics was extended to include quantum system  undergoing quantum measurement, in the absence of any thermal reservoir~\cite{Alonso16,Elouard17role,Benoist18,Dressel18,
Manikandan18,ElouardChapter}. Indeed this situation leads to quantum trajectories of the measured system that are analogous to the stochastic trajectories in phase space of classical stochastic thermodynamics. The equivalent of the first law and the second law have been derived for generic form of measurements~\cite{Elouard17role}, leading to applications such as an engine fueled by the quantum measurement process~\cite{Elouard17extracting,Yi17,Elouard18efficient}.  
In~\cite{Dressel18,Manikandan18}, a new arrow of time measure was introduced to describe the irreversibility of continuous quantum measurement on qubits. Such weak measurements do not completely project the qubit's wavefunction on an eigenstate of the measured observable and therefore generate coherent diffusive trajectories of the state of the measured system. They have been studied intensively \cite{chantasri2013action,Mensky79,Caves87,Wiseman96,Steck06,weber2014mapping,Chantasri15} and provide a wide range of applications exploiting their low invasiveness with respect to strong (projective) measurements \cite{korotkov2006undoing,siddiqi2006dispersive,vijay2012stabilizing,campagne2014observing,campagne2016observing,gillett2010experimental,kim2012protecting,Jordan10}, which justifies to extend quantum stochastic thermodynamics to describe them. The approach followed here relies on the fact that, just as the dynamics of classical systems, continuous measurements on qubit are microscopically reversible and can be undone~\cite{Jordan10,Manikandan18,Dressel18,korotkov2006undoing}, but yet a statistical arrow of time can be identified for the set of quantum trajectories.
\begin{figure}\label{f:setup}
    \includegraphics[width=\linewidth]{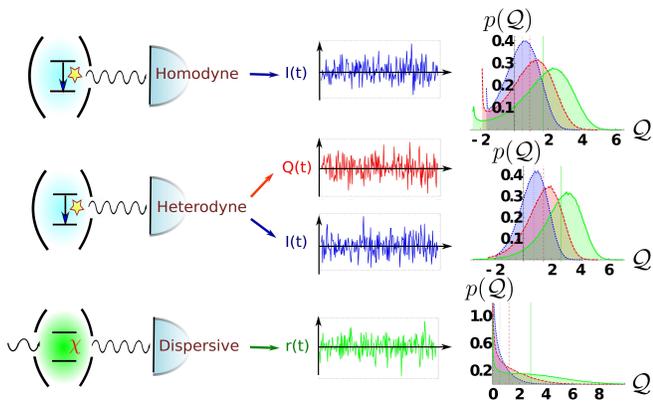}
\caption{Three different continuous measurement schemes compared in the manuscript. Top: Homodyne detection of qubit fluorescence [single readout, $I(t)$]. Middle: Heterodyne detection of qubit fluorescence [two readouts, $Q(t),~I(t)$]. Bottom: dispersive spin measurement, having a single readout $r(t)$. In each case, we plot an example of measurement record (the amplitude is in arbitrary units), and the probability distribution of the arrow of time measure $\Q$ for different measurement durations  $T=0.5\tau$ (blue, dotted), $T=\tau$ (red, dashed), $T = 2\tau$ (green, joined). The qubit is initialized in the eigenstate of $\sigma_x$ with eigenvalue $1$. We have set $\gamma^{-1}= \tau$. The remarkable shape of $P(\Q)$ for the Homodyne (top) and dispersive (bottom) schemes is analytically explained in the SM. \label{f:Setup}}
\end{figure}

While previous studies~\cite{Dressel18,Manikandan18} focused on the average of the defined arrow of time measure, we show here that it is constrained by a FT analogous to those previously derived for the entropy produced in contact with a heat bath. We demonstrate that continuous quantum measurement leads to absolutely irreversible dynamics: just as the free expansion of a gas, the wavefunction collapse generates backward trajectories without forward counterparts. Since we apply a time reversal rule which exactly reverses the quantum state dynamics, and the arrow of time is computed without any projective end point measurement,
the resulting FT with absolute irreversibility is different from its previous appearences~\cite{morikuni2011quantum,Funo15,teifel2010limitations}, and from other quantum generalizations of FTs in general~\cite{Kurchan01,Mukamel03,Campisi09,Esposito09,Campisi11,Chetrite12,Horowitz12,
Horowitz13,campisi2011influence,campisi2010fluctuation,Naghiloo18,Naghiloo17,miller2017time}.
We apply our results for different measurement schemes on a qubit, highlighting how the arrow of time varies in these different contexts, and investigating the influence of measurement strength. 

This Letter is organized as follows: We first introduce the arrow of time measure for a simple two outcome weak measurement of the qubit population, and then for a general weak measurement performed on a qubit. We then express the FT and its proof. Finally, we apply our formal results to several physical systems.

{\it Arrow of time measure}--- We consider a qubit of Hamiltonian $H_0 = (\hbar\omega_0/2)\sigma_z = (\hbar\omega_0/2)(\ket{e}\bra{e}-\ket{g}\bra{g})$, initially in a pure state $\ket{x_0}$ and then weakly measured. In order to introduce our arrow of time measure, we first consider that the measurement is a weak discrete measurement of the qubit population characterized by the two following Kraus operators $M_k(r)$, associated with outcomes $r\in\{1,-1\}$:
\bb\label{eq:Mk}
M_k(1) = \left(\begin{array}{cc} \!\!\sqrt{1-k} &0\\ \!\! 0 &\sqrt{k} \end{array}\!\right),
\; M_k(-1) = \left(\begin{array}{cc} \!\!\sqrt{k} &0\\ \!\! 0 &\sqrt{1-k} \end{array}\!\right).\,
\ee
This POVM models, for example, a weak polarization measurement using a single photon meter~\cite{Pryde04}. The parameter $k \in [0,1/2]$ quantifies the measurement strength ($k=0$ corresponds to a strong measurement, $k = 1/2$ corresponds to a non-informative measurement). After the measurement, the qubit is in state $\ket{x_1(1)}\propto M_k(1)\ket{x_0}$ (resp. $\ket{x_1(-1)}\propto M_k(-1)\ket{x_0}$) when outcomes $r =\pm1$ are obtained. As $M_k(r)M_k(-r)$ is proportional to identity, the forward trajectory $\Gamma_{|x_{0},r} \equiv \{x_0,x_1(r)\}$ is reversed (i.e. the qubit follows the backward trajectory $\tilde\Gamma_{|x_{1}(r),r} \equiv \{x_1(r),x_0\}$) when Kraus operator $M_k(-r)$ is applied on $\ket{x_1(r)}$. This reversal of the measurement is stochastic; it requires the result $-r$ is realized, which occurs with probability $P_B[r\vert x_1(r)] = \|\tilde M_k(r)\ket{x_1(r)}\|^2$, where we have denoted $\tilde M_k(r) = M_k(-r)$ the backward Kraus operator associated with $M_k(r)$. A quantitative measure of the arrow of time can then be obtained by comparing the probabilities $P_F[\Gamma_{|x_{0},r}] = P_F[r\vert x_0]$ and $P_B[\tilde\Gamma_{|x_{1}(r),r}] = P_B[r\vert x_1(r)]$. We define the quantity ${\cal Q}_k(\Gamma_{|x_{0},r}) = \log\{P_F[\Gamma_{|x_{0},r}]/P_B[\tilde\Gamma_{|x_{1}(r),r}]\}$, here given by ${\cal Q}_k(\Gamma_{|x_{0},r})= \log\{[(r+ z_{0}-2kz_{0})^{2}]/[4k(1-k)]\}$, with $z_0 = \bra{x_0}\sigma_z\ket{x_0}$. The sign of ${\cal Q}_k(\Gamma_{|x_{0},r})$ indicates which time-direction of the trajectory -- forward or backward -- is the most probable \cite{Dressel18,Manikandan18}. Note that ${\cal Q}_k(\Gamma_{|x_{0},r})$ diverges in the limit $k\to 0$, which is consistent with the fact that an ideal strong measurement has a zero probability to be reversed this way. Interestingly, the average over the measurement outcomes $\moy{{\cal Q}_k(\Gamma_{|x_{0},r})}_{r}$ is non-negative for any value of $k$ [see supplemental materials (SM)], demonstrating that a clear arrow of time emerges in the measurement process despite microscopic reversibility. 
The initial condition $z_0 = \mp 1$ corresponds to a fixed point of the measurement, leading to deterministic quantum state dynamics independent from the records. Yet, when $k\in[0,1/2)$, one finds a non-vanishing arrow of time reflecting the probabilistic nature of the weak measurement readout $r$.

We now want to study weak measurements with continuous outcomes, performed during some finite time $T = Ndt$ on the qubit. The evolution of the qubit follows a quantum trajectory defined by the set of Kraus operators $\{M(r_n)\}_{0\leq n \leq N-1}$ associated with elementary outcomes $r_n$ obtained at times $t_n = ndt$. We introduce $\r = \{r_0,...,r_{N-1}\}$ the measurement record obtained in a single realization of the process which together with the initial state $x_0$ uniquely defines a quantum trajectory \bb\Gamma_{|x_{0},\r}\equiv\{x_{0},~x_{1}(r_{0}|x_{0}),~x_{2}(r_{1}|x_{1})~...~x_{N}(\r)\},\ee followed by the qubit. We denote $x_{N}(\r)=x_{N}[r_{(N-1)}|x_{(N-1)}]$ for brevity. The probability density of the records reads $P_F(\Gamma_{|x_{0},\r})\equiv P_F(\r\vert x_0) = \|\overleftarrow{\prod_n}M(r_n)\ket{x_0}\|^2$  where the arrow indicates that the operators are ordered from right to left \cite{Chantasri15,Manikandan18}.

It has been demonstrated in~\cite{Manikandan18} that the trajectory $\Gamma_{|x_{0},\r}$ 
followed by the qubit when record $\r$ is obtained can be reversed by applying the Kraus operators given by:
\bb\label{eq:Mtilde}
\tilde M(r_n) = \theta^{-1} M^\dagger(r_n) \theta
\ee
on the final state $\ket{x_N(\r)}$, in reversed order [i.e. starting with $\tilde M(r_{N-1})$]. Here $\theta$ is the time-reversal operator, which in the case of rank-2 Kraus operators ensures $\tilde{M}(r_n)M(r_n) \propto \idop$ \cite{Manikandan18}. Applying $\tilde M(r_n)$ sequentially generates the backward trajectory
$\tilde\Gamma_{|x_{N}(\r),\tilde\r} \equiv\{x_{N}(\r)~...~x_{0}\}$, bringing the qubit through the same sequence of states, in reversed order, back to $|x_{0}\rangle$. The trajectory is reversed with a finite probability $P_B[\tilde\Gamma_{|x_{N}(\r),\tilde\r}]\equiv P_B(\tilde\r|x_N)=\|\overleftarrow{\prod_n}\tilde M(\tilde r_n)\ket{x_N}\|^2$, where $\tilde\r=\{r_{N-n}\}_{1\leq n \leq N}$ is the backward record.  One can then define for any trajectory $\Gamma_{|x_{0},\r}$ the arrow of time measure\bb{\cal Q}(\Gamma_{|x_{0},\r}) = \log\big\{P_F[\Gamma_{|x_{0},\r}]/P^{AC}_B[\tilde\Gamma_{|x_{N}(\r),\tilde\r}]\big\}.\label{Eq:AoT}\ee Here the superscript $\text{AC}$ indicates that we consider the \emph{absolutely continuous part of $P_B$ with respect to $P_F$}, in the sense of Lebesgue's decomposition of probability distributions \cite{halmos2013measure}. In less technical words, $P_B^\text{AC}[\tilde\Gamma_{|x_{N}(\r),\tilde\r}]$ is equal to $P_B[\tilde\Gamma_{|x_{N}(\r),\tilde\r}]$, except when $P_F[\Gamma_{|x_{0},\r}]$ vanishes (when a given backward trajectory does not have a forward counterpart), where it is equal to $0$.

As an example, we review the continuous weak measurement of observable $\sigma_z$, which can be implemented exploiting a dispersive coupling between the qubit and a cavity (see Fig.~\ref{f:Setup}). The evolution of the qubit's state between $t_n$ and $t_{n+1}$ without Rabi drive is  obtained by applying the Kraus operator $M_z(r_n) = (dt/2\pi\tau)^{1/4}e^{-(dt/4\tau)(r_n-\sigma_z)^2}$, with $\tau$ the characteristic measurement time, and $r_n\in\mathbb{R}$. After $T = Ndt$, the qubit's state is $\ket{x_N(\r,x_0)} \propto e^{-(dt/4\tau)\sum_n(r_n-\sigma_z)^2}\ket{x_0}$. The Kraus operators generating the backward dynamics are given by $\tilde M_z(r_n) = M_z(-r_n)$. We obtain the arrow of time in this case, $\mathcal{Q}_{z}$~\cite{Dressel18},
\bb
{\cal Q}_z(\Gamma_{|z_{0},\r})= 2\log\left[\,\cosh\!\left(R \right)+z_0\sinh\!\left(R\right)\,\right]\!,\;\;
\ee
where $R = dt\sum_n r_n/\tau$. When $z_0 = 0$ (i.e. when $\ket{x_0}$ lays on the equator of the Bloch sphere), one finds that ${\cal Q}_z(\Gamma_{|z_{0},\r}) >0$ for any $\r$, leading to a strictly positive average~\cite{Dressel18}. This special case is analogous to the example of free expansion of a single particle gas where the entropy production is always positive, subsequently violating the Jarzynski equality~\cite{Murashita14}. We revisit this example in the SM, and analytically verify the FT presented in this letter. 

We emphasize that despite being based on a similar approach, the present arrow of time measure is distinct from the entropy production as defined in~\cite{Manzano15,Elouard17role,Barra17,ElouardChapter}. This is a direct consequence of the different definition for the time-reversal rule. The present time-reversal choice imposes to reverse exactly the quantum system's sequence of states $\x$ as the measurement record is reversed, while other approaches solely impose to reverse the measurement record. A direct consequence of this tighter constraint is that the present approach is valid solely when the Kraus operators are invertible (i.e. rank-2 when the system is a single qubit). Interestingly, this method leads to an arrow of time measure particularly well-suited for continuous measurement and zero temperature, two limits in which the traditional form of entropy production generally lead to divergences~\cite{Elouard17role,ElouardChapter}. In the remainder of this letter, we show that our arrow of time measure satisfies a FT similar to the Integral Fluctuation Theorem for the entropy production, extensively studied in the case of a quantum system in contact with a thermal reservoir~\cite{Manzano15,Elouard17role,saito2007fluctuation,deffner2011nonequilibrium,yi2013nonequilibrium,Campisi09}. We will apply our general results to four different measurement schemes: the two examples already presented, and the detection of the fluorescence of the qubit via a Heterodyne setup (i.e. after a phase-preserving amplification of the field yielding information on both its quadratures $I_n$ and $Q_n$, stored in the record $r_n = I_n - i Q_n\in\mathbb{C}$) and a Homodyne setup (after a phase-sensitive amplification of the field gathering information about one quadrature stored in $r_n\in\mathbb{R}$)~\cite{Murch13}. The Kraus operators encoding the effect of such measurements during a small time step $dt$ read:
\bb
M_\text{He}(r_n) &=& \frac{e^{-\vert r_n \vert^2/2}}{\sqrt\pi}\left(\begin{array}{cc} \!\!\sqrt{1-\epsilon} &0\\ \!\! \sqrt{\epsilon}\,r_n^* &1 \end{array}\!\right),\nonumber\\
M_\text{Ho}(r_n) &=&\frac{e^{-r_n^2/2}}{\pi^{1/4}}\left(\begin{array}{cc} \!\!\sqrt{1-\epsilon/2} &0\\ \!\! \sqrt{\epsilon}\, r_n  &1 \end{array}\right),
\ee
where $\epsilon = \gamma dt$, with $\gamma$ the spontaneous emission rate of the qubit. The backward evolution operators and the arrow of time measure $\Q$ can be computed following the same protocol described in Eq.~\eqref{eq:Mtilde} and Eq.~\eqref{Eq:AoT}.  Their probability distributions are plotted in Fig.~\ref{f:Setup} for the three different continuous detection schemes, highlighting their strictly positive average value. Interestingly, the average value of the arrow of time measure depends on the measurement scheme, even though the system being measured in these cases is the same, and the measurement rates are chosen to be identical $\gamma = 1/\tau$. We also study the case of continuously monitoring a qubit undergoing Rabi oscillations, in the SM. 

 {\it Fluctuation theorem} --- To obtain our FT, we compute the average value of $e^{-\Q(\Gamma)} = P^{AC}_B(\tilde\Gamma)/P_F(\Gamma)$ over the forward trajectories $\Gamma$, i.e. $\moy{e^{-\Q(\Gamma)}} = \int D\Gamma~P_F(\Gamma
 )~e^{-\Q(\Gamma)}$. Since we need to integrate over all possible realizations, the constraint that the measurement readout $\textbf{r}$ and the quantum state dynamics $\textbf{x}$ at each step correspond via the Bayesian update rule for each individual realizations is imposed by defining $\int D\Gamma$ appropriately as $ \int D\Gamma= \int D\textbf{x}\int D\textbf{r}~\delta[\textbf{x}-\textbf{x}(\textbf{r})]$ (see SM).  We find the central result of this letter:
 \bb
 \moy{e^{-{\cal Q}(\Gamma)}} &=& 1-\mu, \label{eq:IFT}
 \ee
 where $\mu$ is a parameter equal to (see SM):
 \bb\label{eq:muE}
 \mu=1-\int D\Gamma P_B^\text{AC}(\Gamma)= \int D\r \frac{\vert\bra{\bar x_0}{\cal M}^\dagger(\r){\cal M}(\r)\ket{x_0}\vert^2}{\bra{x_0}{\cal M}^\dagger(\r){\cal M}(\r)\ket{x_0}},~~~
 \ee
where ${\cal M}(\r) =\overleftarrow{\prod_n}M(r_n)$ is the global Kraus operator of the sequence of measurements and $\ket{\bar x_0}$ is the normalized state orthogonal to $\ket{x_0}$. From Eq.~\eqref{eq:muE} it is clear that $\mu \geq 0$. The equality $\mu = 0$ can be reached solely if $\ket{x_0}$ is an eigenstate of the global effect operator ${\cal E}(\r) = {\cal M}(\r)^\dagger{\cal M}(\r)$ for any $\r$. Applying the Cauchy-Schwartz equality for vectors $\ket{\psi} = {\cal M}(\r)\ket{x_0}$ and $\ket{\phi}= {\cal M}(\r)\ket{\bar x_0}$ yields $\vert\bra{\phi}\psi\rangle\vert^2/\bra{\psi}\psi\rangle \leq \bra{\phi}\phi\rangle$, which demonstrates that $\int D\r \bra{\bar x_0}{\cal M}^\dagger(\r){\cal M}(\r)\ket{\bar x_0} = 1$ is an upper bound for $\mu$. 
 
Equality \eqref{eq:IFT} constrains the fluctuations and average of the arrow of time measure. In particular, it readily imposes via Jensen's inequality a lower bound on the average arrow of time:
 \bb
 \moy{{\cal Q}(\Gamma)} &\geq & - \log(1-\mu)\label{eq:2ndLaw}.
 \ee
 
 {\it Absolute irreversibility} --- The r.h.s of the FT in Eq.~\eqref{eq:IFT} is strictly lower than $1$ when the initial state is not an eigenstate of the effect matrix, leading to a strictly positive value of $\moy{\Q(\Gamma)}$. This feature has been referred to as absolute irreversibility~\cite{Murashita14,Funo15}, and reveals existence of time-reversed trajectories that are accounted for by probability law $P_B$, but which do not bring the system back to its initial state $\ket{x_0}$. For such trajectories, the forward probability is zero so that the ratio $P_B(\tilde\r \vert x_N)/P_F(\r\vert x_0)$ and the arrow of time diverges~\cite{Funo15}. Taking the absolutely continuous part $P_B^\text{AC}$ of $P_B$ in the definition of $\Q(\Gamma)$ is required to restrict the average in Eq.~\eqref{eq:IFT} to allowed forward trajectories. Though, the existence of backward paths without forward counterpart still play a role in the properties of $\Q(\Gamma)$ by giving a strictly positive value to $\mu$. Technically, one can understand why $\mu$ is non-zero by noting that the integrand in Eq.~\eqref{eq:muE} is not a normalized probability distribution for $\Gamma$. Whereas $P_B(\tilde\r\vert x_f)$ for a fixed $x_f$ is normalized to $1$ when summing over $\tilde\r$, $P_B(\tilde\r\vert x_N)$ also depends on $\x$ through $x_N$, which causes the integral to differ from unity.

Physically, this absolute irreversibility disappears solely when the measurement has no effect on the qubit's state. This situation can still lead to a non-zero $\langle\mathcal{Q}\rangle$ if the measurement outcome fluctuates, for example when applying the measurement operators in Eq.~\eqref{eq:Mk} to an eigenstate of $\sigma_z$. A perfectly reversible situation ($\langle\mathcal{Q}\rangle = 0$) requires in addition that the measurement outcome is certain.
This illustrates that irreversibility ($\langle\mathcal{Q}\rangle>0$) and absolute irreversibility ($\mu\neq 0$) are two different properties defined for a set of forward trajectories which help characterizing the arrow of time in a microscopically reversible process.

We finally emphasize that one can generalize Eq.~\eqref{eq:IFT} to the case where the initial state of the system is drawn from an ensemble $\{\ket{x_0}\}$ with probability $p(x_0)$. This situation still leads to absolute irreversibility in general (see SM).

\begin{figure}
	\includegraphics[width=\linewidth]{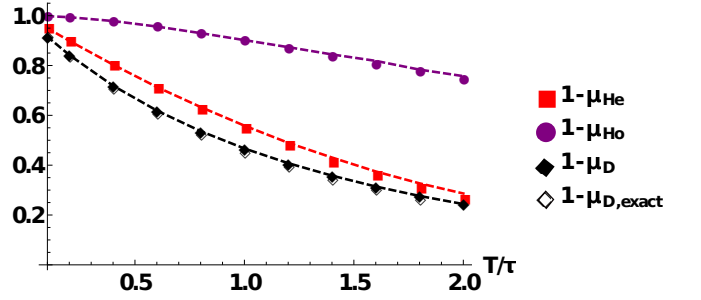}
	\caption{Absolute irreversibility of the three studied continuous detection schemes: Left-hand side of the FT $\moy{e^{-\Q(\Gamma)}}$ (dashed) and parameter $\mu$ (dotted) computed from Eq.~\eqref{eq:muE}, as a function of the duration of the measurement $T/\tau$, starting from $T/\tau=0.1$.  The qubit is initialized in the eigenstate of $\sigma_x$ with eigenvalue $1$. We have simulated $1\times 10^{6}$ trajectories, setting $\tau^{-1}=\gamma$. The analytically obtained value $1-\mu_{\text{D,exact}}$ for the dispersive measurement with no Rabi drive is also marked in the figure.\label{f:FT}}
\end{figure}

{\it Analysis of the examples} --- We first apply our results to the weak measurement characterized by $M_k(\pm 1)$ defined in Eq.~\eqref{eq:Mk}. Here the parameter $\mu$ can be computed analytically:
 \bb
 \mu_k = \big[(1-2k)^2(1-z_0^2)\big]/\big[1-(1-2k)^2z_0^2\big],
 \ee
 which for $k\in[0,1/2]$ indeed belongs to $[0,1]$. We retrieve in this example that $\mu_k = 0$ for $z_0 = \pm1$ and $\mu_k$ is strictly positive otherwise. The limit $k\to 0$ (strong measurement) corresponds to $\mu_k\to 1^-$, such that the bound $-\log(1-\mu_k)$ goes to $+\infty$, capturing that the arrow of time measure diverges for a strong measurement. Conversely, for $k\to 1/2$, $\mu_k$ goes to $0$ for any value of $z_0$: the measurement in this limit does not gather any information and has no effect of the qubit, such that the process becomes absolutely reversible, and $\langle\Q_{k}\rangle\rightarrow 0$. Interestingly, for a fixed $z_0 \in [-1,1]$, the parameter $k$ allows to go from a perfectly strong measurement to a weak measurement, and even to no measurement at all. This transition is accompanied by $\langle\Q_{k}\rangle$ going from $+\infty$ to $0$, and absolute irreversibility is present but its amount, quantified by $\mu_k$ decreases and finally reaches $0$ when the measurement has no back-action anymore on the qubit's state.

For the dispersive $\sigma_{z}$, Homodyne and Heterodyne measurements on a qubit for a finite duration $T=Ndt$, we verify the FT by simulating both a fair sample of qubit trajectories to compute $\moy{e^{-\Q(\Gamma)}}$ and numerically integrate Eq.~\eqref{eq:muE} as shown in Fig.~\ref{f:FT}. One can see the agreement between both sides of Eq.~\eqref{eq:IFT}, which numerically validate our FT, and proves the presence of absolute irreversibility as well as $\mu$ is greater than zero.  We also compare our results to the analytical solution for $\mu$ for the dispersive measurement with no Rabi drive, discussed in the SM. Just as parameter $k$ in the two-outcome measurement example, the measurement time allows to switch between an extremely weak measurement (for $T\ll \tau$) such that $\mu\ll 1$ and $\moy{\Q(\Gamma)}\geq 0$ to a strong measurement (for $T\gg \tau$) such that $\mu$ goes to $1$ and the lower bound for the average arrow of time diverges. The agreement to our FT for single step measurements, and for continuously monitoring a qubit undergoing Rabi oscillations are also presented in the SM.

{\it Conclusion} --- We have proved that the arrow of time measure for continuous measurement on qubits fulfills a fluctuation theorem, just like the entropy production associated with a transformation of a quantum system in contact with a thermal reservoir. This FT allowed us to show that weak continuous quantum measurement exhibits absolute irreversibility, and therefore is associated with a strictly positive average arrow of time measure. A zero lower bound for the average arrow of time is possible only when the qubit is in an eigenstate of the effect matrix of the sequence of measurements. We have analyzed different measurement schemes, including dispersive measurement of a qubit observable, and homodyne and heterodyne measurements of the fluorescence, highlighting how the arrow of time value, and the degree of absolute irreversibility, depends on the chosen type of measurement. This study emphasizes that absolute irreversibility is inherent to the quantum measurement process. Moreover, it paves the road towards a complete thermodynamic description of quantum measurements. Due to the growing importance of schemes based on continuous monitoring in various applications, ranging from metrology to quantum computing or tomography, this is an essential step towards a full understanding of the resource needed to perform useful quantum tasks.
 
 {\it Acknowledgements} -- This work was supported by the John Templeton Foundation Grant ID 58558, the US Army Research Office grant No. W911NF-15-1-0496, the National Science Foundation grants No. DMR-1506081 and NSF PHY-1748958, and the US Department of Energy grant No. DE-SC0017890. C.E. and A.N.J thank Chapman University and the Institute for Quantum Studies for hospitality during this project. We warmly thank Alexander Korotkov, Justin Dressel, Michele Campisi, Alexia Auff\`eves, Massimiliano Esposito, Janet Anders and the other participants to the KITP program QTHERMO18 for helpful discussions.
\bibliography{biblio}
\pagebreak
\widetext
\begin{center}
\textbf{\large Supplemental Materials: Fluctuation Theorems for Continuous Quantum Measurement and Absolute Irreversibility }
\end{center}

\setcounter{equation}{0}
\setcounter{figure}{0}
\setcounter{table}{0}
\setcounter{page}{1}
\makeatletter
\renewcommand{\theequation}{S\arabic{equation}}
\renewcommand{\thefigure}{S\arabic{figure}}

\section{A. Average value of $\Q_{k}$ for the two outcome spin measurement}

\begin{figure}	\includegraphics[width=\linewidth]{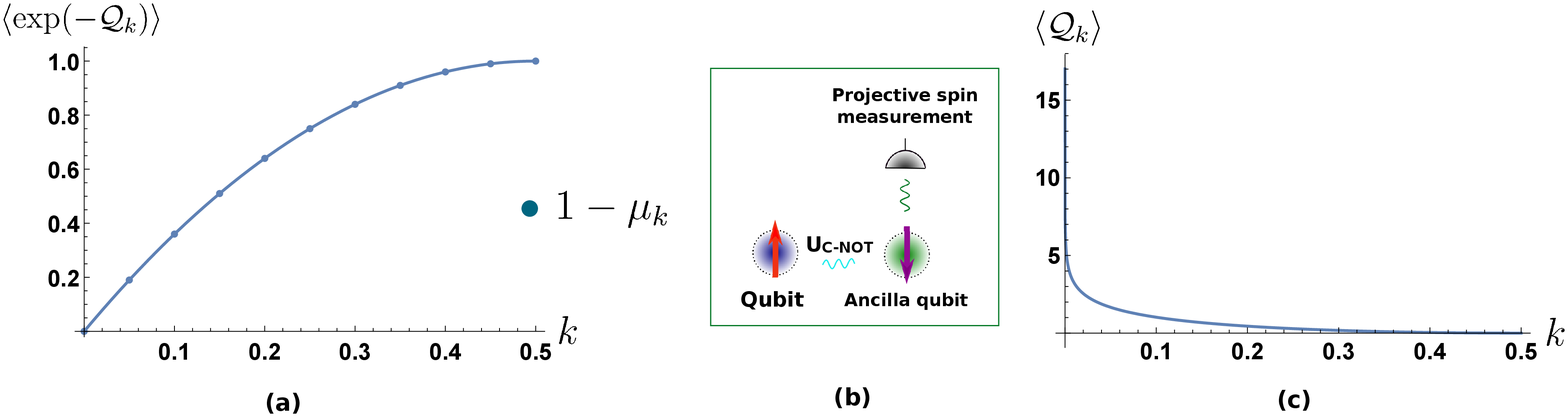}
	\caption{Here we consider a single step weak discrete measurement of qubit population, when the qubit initialized at  $x=1$. In Fig.~\ref{figDis}~(a), we show that the identity $\langle \exp(-Q_{k})\rangle$ [solid line] $=1-\mu_{k}$ [dotted] is satisfied for different values of the measurement strength $k~\epsilon~[0,~\frac{1}{2}]$. A possible experimental implementation of this measurement scheme is shown in Fig.~\ref{figDis}~(b), where the quantum system (qubit) and the measuring device (ancilla qubit) evolve via the controlled-NOT unitary. The measurement is completed by projecting the ancilla qubit onto the spin basis. (c) Here we plot the average value $\langle\Q_{k}\rangle$ for $k~\in~[0,1/2]$ for a qubit initialized at $z=0$, considering the two outcome $z$ measurement discussed in the main text.\label{figDis}}
\end{figure}

For the single step, two outcome spin measurement described by measurement operators,
\bb\label{eq:Mk1}
M_k(1) = \left(\begin{array}{cc} \!\!\sqrt{1-k} &0\\ \!\! 0 &\sqrt{k} \end{array}\!\right),
\; M_k(-1) = \left(\begin{array}{cc} \!\!\sqrt{k} &0\\ \!\! 0 &\sqrt{1-k} \end{array}\!\right).\,
\ee
we compute the average value of $\Q_{k}(\Gamma)$ as $\langle \Q_{k}(\Gamma)\rangle = P_{F}(\Gamma_{1})\Q_{k}(\Gamma_{1})+P_{F}(\Gamma_{-1})\Q_{k}(\Gamma_{-1})$, where $\Q_{k}(\Gamma_{r})$ is computed using the formula  ${\cal Q}_k(\Gamma_{r})= \log\{[(r+ z_{0}-2kz_{0})^{2}]/[4k(1-k)]\}$, for $r\in \{-1,1\}$. In Fig.~\ref{figDis}, we plot the average $\langle\Q_{k}\rangle$ for the case $z_{0} = 0$, that demonstrate the essential features discussed in the main text, its non-negativity, and positive divergence as $k\rightarrow 0.$

\section{B. Derivation of the fluctuation theorem}
Here we derive the identity $\langle e^{-\Q(\Gamma)}\rangle = 1-\mu$, by considering discrete state update using Kraus operators and then taking the continuum limit. We first note that the probability distribution function of the forward state update for a sequence of N measurements -- that imposes the constraint that a given pair $\Gamma=(\x,\r)$ has a non-vanishing probability if and only if the sequence of states  $\textbf{x} = \{x_{k}\}_{k=0}^{N}$ and the measurement readouts $\textbf{r} = \{r_{k}\}_{k=0}^{N-1}$ correspond via the Bayesian update rule: $\x(\r)=\{x_{0},~x_{1}(r_{0}|x_{0}),~x_{2}(r_{1}|x_{1})~...~x_{N}(r_{(N-1)}|x_{(N-1)})\}$ -- can be written as follows~\cite{Chantasri15,chantasri2013action}:
\begin{equation}
\mathcal{P}_{F}(\Gamma) = \delta(x_{0}-x_{in}) \prod_{k=0}^{N-1}P_{F}(x_{k+1}|x_{k},r_{k})P_{F}(r_{k}|x_{k}).\label{eq1}
\end{equation}		
Here the term $P_{F}(x_{k+1}|x_{k},r_{k})$ represents a deterministic state update given the dynamics, imposed as a 3 dimensional $\delta$ function for each component of spin along the Bloch sphere coordinates, 
\begin{equation}
P_{F}(x_{k+1}|x_{k},r_{k})=\prod_{i=1}^{3}\delta\bigg[x_{k+1}^{i}-\text{Tr}\bigg(\hat{\sigma}^{i}\frac{U_{k}M(r_{k})\rho_{k}M(r_{k})^{\dagger}U_{k}^{\dagger}}{\text{Tr}[M(r_{k})\rho_{k}M(r_{k})^{\dagger}]}\bigg)\bigg] =\delta[x_{k+1}-(x_{k+1}|x_{k},r_{k})],
\end{equation}
and the probability of obtaining a readout $r_{k}$ given $x_{k}$ is given by the expression,
\begin{equation}
P_{F}(r_{k}|x_{k}) = \text{Tr}[M(r_{k})\rho_{k}M(r_{k})^{\dagger}].
\end{equation}		
Note that imposing a delta function boundary condition at each step as in Eq.~\eqref{eq1} ensures that the trajectories where ${r_{k}}$ and ${x_{k}}$ do not correspond to each other have probability zero. These trajectories -- completely determined by the initial state $x_{0}$ and the measurement readout $\r$ -- are labeled by the notation $\Gamma_{|x_{0},\r}$ in the main text, referring to individual realizations of the measurement process.

For any given final state $x_{f}$ obtained at the end of the forward measurement, the backward probability distribution can be written similarly,
\begin{equation}
\mathcal{P}_{B}(\tilde\Gamma) =\delta(x_{N}-x_{f}) \prod_{k=N}^{1}P_{B}(x_{k-1}|x_{k}, r_{k-1})P_{B}( r_{k-1}|x_{k})\label{eqb},
\end{equation}	
where we have
\begin{equation}
P_{B}(x_{k-1}|x_{k},r_{k-1})=\prod_{i=1}^{3}\delta\bigg[x_{k-1}^{i}-\text{Tr}\bigg(\hat{\sigma}^{i}\frac{\tilde M(r_{k-1})U_{k-1}^{\dagger}\rho_{k}U_{k-1}\tilde M(r_{k-1})^{\dagger}}{\text{Tr}[\tilde M(r_{k-1})U_{k-1}^{\dagger}\rho_{k}U_{k-1}\tilde M(r_{k-1})^{\dagger}]}\bigg)\bigg] = \delta[x_{k-1}-(x_{k-1}|x_{k},r_{k-1})].
\end{equation}
The update operator $\tilde M(r_{k})=\theta^{-1} M(r_{k})^{\dagger}\theta$, where $\theta$ is the time reversal operator, and the backward probabilities,
\begin{equation}
P_{B}(r_{k-1}|x_{k}) = \text{Tr}[\tilde M(r_{k-1})U_{k-1}^{\dagger}\rho_{k}U_{k-1}\tilde M(r_{k-1})^{\dagger}].
\end{equation}
We now proceed to compute the quantity $\langle e^{-\Q(\Gamma)}\rangle$ as a statistical average over all possible forward trajectories in the ensemble being considered. The integration measure over all the possible trajectories $\Gamma$ with non-vanishing forward probabilities can also be expressed in terms of the readouts $\textbf{r}$ and the corresponding Bloch sphere coordinates $\x$ as, 
\begin{equation}\int D\Gamma = \int D\textbf{x}\int D\textbf{r}~\delta[\textbf{x}-\textbf{x}(\textbf{r})],\end{equation}
where we assume $\int D\textbf{x} \equiv \int\prod_{k=1}^{N} Dx_{k}$. Note that the Bloch sphere coordinates $x_{k}$ take continuum of values in the interval $[-1,1]$, and the readout(s) \textbf{r} for the Homodyne/ Heterodyne measurements are also continuous variables. The $\delta$ function imposes the constraints of the initial state and the Bayesian state update,
\begin{equation}
\delta[\textbf{x}-\textbf{x}(\textbf{r})] = \delta(x_{0}-x_{in})\prod_{k=0}^{N-1}\delta[x_{k+1}-(x_{k+1}|x_{k},r_{k})].\end{equation} 
The quantity $\langle e^{-\Q(\Gamma)}\rangle$ pertinent to our time-reversal scheme is defined as the following integral over paths:
\begin{equation}
\langle e^{-\Q(\Gamma)}\rangle = \int D\Gamma~P_{F}[\Gamma]~\frac{P_{B}^{AC}[\Gamma]}{P_{F}[\Gamma]}.
\end{equation} 
Here for a given trajectory $\Gamma$, we have defined $P_{F}[\Gamma] = \prod_{k=0}^{N-1}P_{F}(r_{k}|x_{k})$. We have also defined $\mathcal{Q} = \log\frac{P_{F}[\Gamma]}{P^{AC}_{B}[\Gamma]}$, where $P^{AC}_{B}[\Gamma]$ correspond to the probability of obtaining a backward trajectory which has a corresponding forward trajectory (having forward probability $P_{F}[\Gamma]$) in the ensemble of all forward trajectories (denoted by the superscript $AC$, implying absolute continuous part of the backward distribution, relative to the forward distribution, used in the context of Lebesgue's decomposition theorem~\cite{halmos2013measure}). This probability of obtaining a readout backward, given the intital state state $x_{0}$ and measurement record $\textbf{r}$ can be written more concisely in terms of the effect matrix as,
\begin{equation}
P^{AC}_{B}[\Gamma]= \prod_{k=N}^{1}P_{B}(r_{k-1}|x_{k}) = \frac{\text{Det}[\cal E(\textbf{r})]}{\text{Tr}[\rho_{x_{0}}\mathcal{E}(\textbf{r})]}.
\end{equation}
Using Eq.~\eqref{eq1} we have,
\begin{eqnarray}
\langle e^{-\Q(\Gamma)}\rangle &=& \int D\Gamma~P_{F}[\Gamma]~\frac{P^{AC}_{B}[\Gamma]}{P_{F}[\Gamma]}=\int D\textbf{x}\int D\textbf{r}~ \mathcal{P}_{F}\frac{\prod_{k=N}^{1}P_{B}(r_{k-1}|x_{k})}{\prod_{k=0}^{N-1}P_{F}(r_{k}|x_{k})}\\&=&\int D\textbf{x}\int D\textbf{r}~\delta(x_{0}-x_{in})\prod_{k=0}^{N-1}\delta[x_{k+1}-(x_{k+1}|x_{k},r_{k})]\prod_{k=N}^{1}P_{B}(r_{k-1}|x_{k})\nonumber\\
&=&\int D\textbf{x}\int D\textbf{r}~\delta[\textbf{x}-\textbf{x}(\textbf{r})]~\frac{\text{Det}[\cal E(\textbf{r})]}{\text{Tr}[\rho_{x_{0}}\cal E(\textbf{r})]} = \int D\textbf{r}~\frac{\text{Det}[\cal E(\textbf{r})]}{\text{Tr}[\rho_{x_{0}}\cal E(\textbf{r})]}.
\end{eqnarray}
\begin{figure}
\includegraphics[scale=0.65]{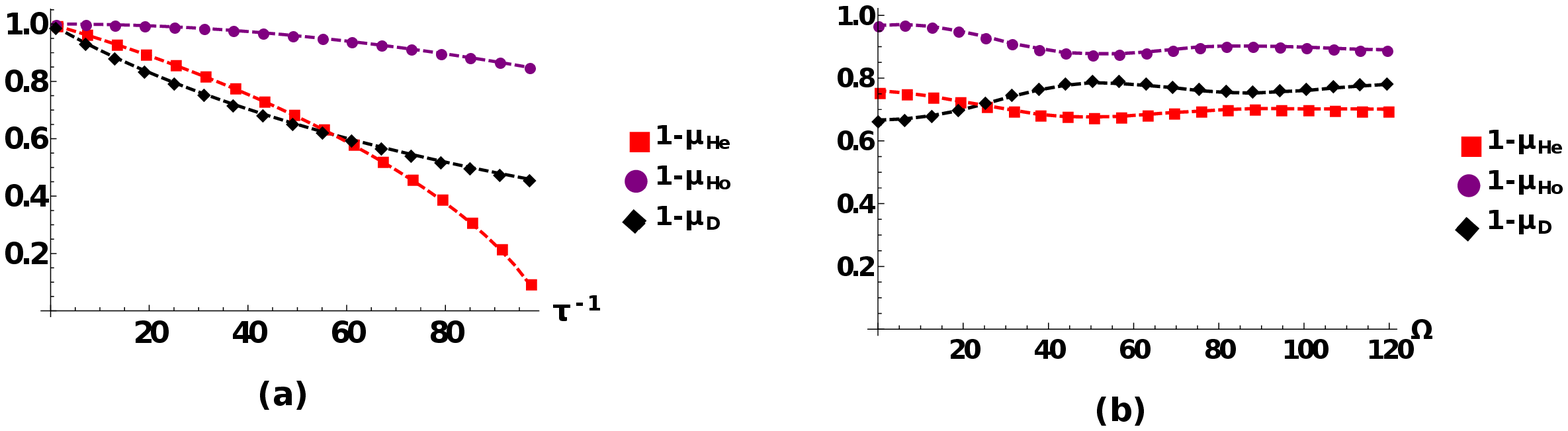}
	\caption{Absolute irreversibility of the three studied continuous detection schemes for a single step measurement: Left-hand side of the FT $\moy{e^{-\Q(\Gamma)}}$ (dashed) and parameter $\mu$ (dotted) computed from Eq.~\eqref{eqmu}, as a function of the duration of the measurement rate $\tau^{-1}=\gamma$.  The qubit is initialized in the eigenstate of $\sigma_x$ with eigenvalue $1$. (b) Verifying the FT for the three studied continuous detection schemes for different Rabi drive frequency $\Omega$: Left-hand side of the FT $\moy{e^{-\Q(\Gamma)}}$ (dashed) and parameter $\mu$ (dotted) computed from Eq.~\eqref{eqmu} for $T=0.5\tau$.  The qubit is initialized in the eigenstate of $\sigma_x$ with eigenvalue $1$. We have simulated $1\times 10^{6}$ trajectories, setting $\tau^{-1}=\gamma$.\label{ss}}
\end{figure}
We performed the integration over \textbf{x} since the integrant depends only on $\textbf{r}$ and $x_{0}$. We now write the effect matrix $\cal E(\textbf{r})$ in the basis of $\{|x_{0}\rangle,~|\bar{x}_{0}\rangle\}$, where $\rho_{x_{0}}=|x_{0}\rangle\langle x_{0}|$, and $\langle x_{0}|\bar{x}_{0}\rangle$ = 0 as:
\begin{equation}
\cal E(\textbf{r})=\begin{bmatrix}
a(\textbf{r})&c(\textbf{r})\\c^{*}(\textbf{r})&b(\textbf{r})
\end{bmatrix}.
\end{equation}
For a given intial state, sum over all probabilities in the forward direction is equal to one implies that the effect matrix $\cal E(\textbf{r})$ satisfies the following relation:
  \begin{equation}
  \int D\textbf{r}~\cal E(\textbf{r}) = 
  \begin{bmatrix}
  1&0\\0&1
  \end{bmatrix}.
  \end{equation}
We therefore obtain,
\begin{eqnarray}
\langle e^{-\Q(\Gamma)}\rangle
&=&\int D\textbf{r}~\frac{\text{Det}[\cal E(\textbf{r})]}{\text{Tr}[\rho_{x_{0}}\cal E(\textbf{r})]}=\int D\textbf{r}~\frac{a(\textbf{r})b(\textbf{r})-|c(\textbf{r})|^{2}}{a(\textbf{r})}\nonumber\\
&=&\int D\textbf{r}~b(\textbf{r})-\int D\textbf{r}\frac{|c(\textbf{r})|^{2}}{a(\textbf{r})}=1-\mu\label{eqmu},
\end{eqnarray}
where we have defined,
\begin{equation}
\int D\textbf{r}~\frac{|c(\textbf{r})|^{2}}{a(\textbf{r})} \equiv \mu,\label{Eq:lamu}
\end{equation}
leading to Eq.~(7) of the main text. We verify this identity in Fig.~\ref{ss}, considering (a) single step measurement described by measurement operator $M_{X}$, and (b) continuously monitoring a qubit subject to Rabi drive, where the effective time evolution operator is $\mathcal{U}(r_{n},dt) = M_X(r_{n})~e^{-\frac{i}{\hbar}Hdt}$ (for $H=\hbar\Omega\sigma_{y}/2$), with $X= z,~ \text{He},~\text{Ho}$, labeling continuous dispersive $\sigma_{z}$ measurement, Heterodyne and Homodyne detection of qubit's fluorescence respectively.  Eq.~\eqref{eqmu} can be analytically verified in certain special cases. An example of such a case is presented in Sec.~D, where we look at the dispersive spin measurement with no Rabi drive, and obtain a probability distribution that estimates $\mu$ analytically.

\section{C. FT in the case of a random initial qubit state}

We now assume that the initial state of the system is drawn from a set $\{\ket{x_0}\}$ according to a probability law $p(x_0)$. As the consequence, the average over the trajectory involved in the fluctuation theorem Eq.~\eqref{eqmu} now corresponds to  $\moy{\cdot} = \int dx_0 p(x_0)\int D\Gamma_{|x_{0}}P_F[\Gamma](\cdot)$ instead of $\moy{\cdot}_{|x_{0}} = \int D\Gamma_{|x_{0}}P_F[\Gamma](\cdot)$ we used earlier, although we had suppressed the conditioning on $x_{0}$ for brevity in our earlier discussions [and in Eq.~(7) of the main text], by absorbing it to the delta function constraint involved in the integration measure $\int D\Gamma$. On the other hand, the definition of the arrow of time measure $\Q(\Gamma_{|x_0,\r})$ associated with a given initial state $x_0$ and record $\r$ is unchanged. We emphasize that the sum over $x_0$ runs onto the qubit's Hilbert space, and the distribution $p(x_0)$ is allowed to be either discrete (e.g. when the preparation is due to the projective measurement of an observable) or continuous.

In this situation, the IFT becomes:

\bb
 \moy{e^{-{\cal Q}(\Gamma)}} &=& 1-\moy{\mu}_{x_0},
 \label{eq:IFTQH}
 \ee
 where $\moy{\mu}_{x_0}$ is the average of $\mu$ over the distribution of initial state:
 \bb
 \moy{\mu}_{x_0} = \int dx_0 p(x_0)\mu = \int dx_0p(x_0)\int D\r \frac{\vert\bra{\bar x_0}{\cal M}^\dagger(\r){\cal M}(\r)\ket{x_0}\vert^2}{\bra{x_0}{\cal M}^\dagger(\r){\cal M}(\r)\ket{x_0}}.
 \ee
 
 In general, such average is not a sufficient condition to have $\moy{\mu}_{x_0}=0$, even when drawing the state from a set of states preserved by the measurement. A simple example is the case of the two-outcome spin measurement described by Eq.~\eqref{eq:Mk1}, applied to a state drawn from the circle of the qubit states of zero $y$ coordinate in the Bloch sphere. One gets:
 \bb
 \moy{\mu}_{x_0} = \int_{-1}^1 dz_0 p(z_0)\frac{(1-2k)^2(1-z_0^2)}{1-(1-2k)^2z_0^2},
 \ee
which takes for instance the value $1-4k(1-k)\text{ArcTanh}(1-2k)/(1-2k)\neq 0$ for a flat probability distribution  $p(z_0) = 1/2$ of the initial $z$ coordinate denoted $z_0$.

This contrast with usual FTs with absolute irreversibility is explained by our choice of (i) defining the arrow of time from the probabilities of the forward (resp. backward) trajectory, conditioned to the initial (resp. final) state, rather than from a joint probability $p(x_0)P_F[\Gamma_{|x_0,\r}]$ (resp. $p(x_N)P_B^\text{AC}[\tilde\Gamma_{|x_N,\tilde\r}]$) of picking the initial (resp. final state) and obtaining the record $\r$; and (ii) not performing a final projective measurement on the system. If one adds these two conditions, one finds another fluctuation theorem of the form:
\bb
\moy{e^{-\Q(\Gamma) - \Delta s[\Gamma]} = 1-\mu'}.
\ee
Here $\Delta s[\Gamma] = \log[p(x_0)/p(x_N)]$ is a boundary contribution that corresponds to the difference of stochastic entropies of the initial and final set of qubit states. In this case, the absolute irreversibility parameter $\mu'$ vanishes provided $p(x_0)$ has a support spanning every final states of the reversed trajectories. The price to pay is that the fluctuation theorem does not involve only the arrow of time measure, but also $\Delta s[\Gamma]$. In addition, one can expect that the final projective measurement has a strong impact, possibly overcoming the contribution of the weak continuous measurement under study.

\section{D. Special case: Dispersive measurement with no Rabi drive\label{Sec. NR}}
\begin{figure}	\includegraphics[width=\linewidth]{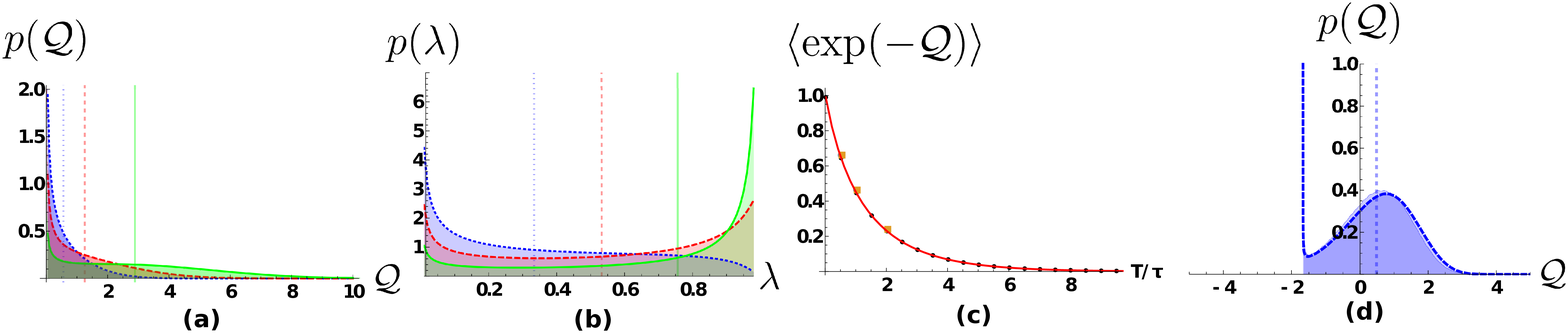}
	\caption{Here we plot (a) the distribution of $\mathcal{Q}$~\cite{Dressel18} indicating their strictly positive average value, and (b) the distribution of $\lambda$, indicating their mean value $\langle\lambda\rangle = \mu$, for different durations: $T/\tau=0.5$ (dotted, blue), $T/\tau=1$ (dashed, red) and $T/\tau=2$ (joined, green), and compare with the numerical simulation of $10^{6}$ trajectories in each case. (c) We verify the fluctuation theorem for dispersive qubit measurement with no Rabi drive starting at $z_{0}=0$ for different values of $T/\tau$.   Left-hand side of the FT $\moy{e^{-\Q}}$ (solid) and parameter $1-\mu$ (dotted) computed from Eq.~\eqref{eqmu}, using the analytical approach discussed in Sec.~D. The data obtained using numerical simulations used in (a) and (b) are indicated using (overlapping) blue circle and orange square markers. (d) Here we compare the analytical solution obtained in Sec. E with the numerical simulation of $10^{6}$ trajectories for $\epsilon' =T/\tau = 0.5$.\label{nR}}
\end{figure}
 Here we look at the particular case of dispersive measurement with no Rabi drive, where the total integrated signal $R=\frac{1}{\tau}\int_{0}^{T}dt~r(t)$ completely describes the measurement dynamics. The probability distribution of $\mathcal{Q}$ in this case can be obtained by methods described in~\cite{Dressel18}, that allows us to compute $\langle \exp(-\Q)\rangle$ analytically as the integral $\langle \exp(-\Q)\rangle = \int d\Q\exp(-Q)\mathcal{P}(\Q)$. Here we note that a similar analytical result can be obtained for $\mu$ as well, that permits us to analytically verify the identity $\langle \exp(-\Q)\rangle=1-\mu$. In order to achieve this, we define $\mu$ as the mean value of the probability distribution of a random variable $\lambda$ 
 \begin{equation}\lambda(R) =\frac{\text{Tr}[\rho(0)\mathcal{E}(R)\tilde{\rho}(0)\mathcal{E}(R)]}{\text{Tr}[\rho(0)\mathcal{E}(R)]^2}.\end{equation}
 Note that $\lambda = \frac{|c(R)|^{2}}{a(R)^{2}}$, by multiplying and dividing the integrand of the l.h.s. of Eq.~\eqref{Eq:lamu} by the forward probability $a(R)$. Here $\rho(0)$ is the initial state, which is assumed to be pure, and $\tilde{\rho}(0)$ is the state orthogonal to that. The probability distribution $\mathcal{P}(\lambda)$ can be obtained from the probability distribution $\mathcal{P}(\Q)$ by noting that,
 \begin{equation}
 \mathcal{P}(\Q)~d\Q = \mathcal{P}(\lambda)d\lambda,
 \end{equation} 
 or,
 \begin{equation}
\mathcal{P}(\lambda) = \mathcal{P}(\Q)\frac{d\Q}{d\lambda}\bigg|_{\Q= \Q(\lambda)}.
 \end{equation}
 We note that for the case when qubit is initialized at $z=0$, this result is rather simple. In this case, we obtain $\lambda(R) = (\tanh R)^{2} = 1-\exp(-\Q)$, where $\Q = 2\log\cosh R$ for the initial state $z = 0$ , as obtained in~\cite{Dressel18}. We obtain,
 \begin{equation}
 \frac{d\Q}{d\lambda} = \frac{1}{1-\lambda}.
 \end{equation}
 Using the relation $\Q(\lambda) = -\log(1-\lambda)$, we obtain the following expression for $\mathcal{P}(\lambda)$ (for qubit initialized at $z=0$),
 \begin{equation}
 \mathcal{P}(\lambda) = \sqrt{\frac{\tau}{2\pi T}}\frac{1}{(1-\lambda)^{2}}\sqrt{\frac{1-\lambda}{\lambda}}\exp\bigg(-\frac{T}{2\tau}-\frac{\tau}{2T}\bigg[\text{arccosh}\frac{1}{\sqrt{1-\lambda}}\bigg]^{2}\bigg)\hspace{1cm}\lambda~\epsilon~[0,~1].
 \end{equation}
 We note that $\mu = \langle\lambda\rangle = \int_{0}^{1}d\lambda~\lambda~\mathcal{P}(\lambda),$ that satisfies $\langle \exp(-\Q)\rangle=1-\mu$. Please refer to Fig.~\ref{nR} where we numerically verify this identity for different durations of the measurement $T/\tau$. \\
 
 \section{E. Homodyne measurement}
 
 From the Kraus operator $M_\text{Ho}$ given in main text, we first compute the arrow of time measure corresponding to a single step homodyne measurement performed during $dt$. We use the identity
 \bb
 \Q(\r) = -\log\left(\dfrac{\vert\text{Det}[M(r)]\vert^2}{\text{Tr}\{\rho_{x_0}M^\dagger(r)M(r)\}^2}\right).
 \ee
 We find for $x_0$ being the eigenstate of $\sigma_x$ of eigenvalue $+1$:
  \bb\label{SM:QHodt}
 \Q_\text{Ho}(r) = \log\left(\frac{1-\epsilon/4+\sqrt{\epsilon}r +\epsilon r^2/2}{1-\epsilon/2}\right).
 \ee
 This expression allows to check that, $\Q_\text{Ho}(\r)$ admits a minimum negative value $\Q_\text{min} =2\log[\sqrt{1-\epsilon/2}/2]$, reached for $r_\text{min}=-1/\sqrt\epsilon$. The probability $P_\text{Ho}^{(dt)}(\Q)$ for $\Q_\text{Ho}$ to take the value $\Q$ is given by:
 \bb
P_\text{Ho}^{(dt)}(\Q) = \left.P(r\vert x_0) \left(\frac{d\Q_\text{Ho}(r)}{dr}\right)^{-1}\right\vert_{r=r(\Q)},
 \ee
 with 
 \bb
 P(r\vert x_0) = \frac{e^{-r^2}}{\sqrt{\pi}}\left(1+\sqrt{\epsilon}r -\frac{\epsilon}{4}+\frac{r^2\epsilon}{2}\right)
 \ee
 and $r(\Q)$ is obtained inverting Eq.~\eqref{SM:QHodt}:
 \bb
 r(\Q) = \frac{1}{\sqrt{\epsilon}}\left(\sqrt{e^{\Q/2}\sqrt{1-\frac{\epsilon}{2}}+\frac{\epsilon}{2}-1}-1\right).
 \ee
 For finite durations of the measurement, the concatenated measurement operators can be written as a single effective measurement,\bb
  {\cal M}_\text{Ho}(r) = \frac{e^{-\sum_{n=1}^N r_n^2/2}}{\pi^{N/4}}\left(
  \begin{array}{cc}
  (1-\epsilon/2)^{N/2} & 0\\
  \sqrt{N\epsilon}~y(\r) & 1
  \end{array}\right)\simeq \frac{e^{-\sum_{n=1}^N r_n^2/2}}{\pi^{N/4}}\left(
  \begin{array}{cc}
  \sqrt{1-\epsilon'/2} & 0\\
  \sqrt{\epsilon'}~y(\r) & 1
  \end{array}\right).
  \ee
  with the effective readout $y(\r)=\frac{1}{\sqrt{N}}\sum_{n=1}^N r_n(1-\epsilon/2)^{(n-1)/2}$, and $\epsilon'=N\epsilon$, and this approximation is valid when $\epsilon\ll1$. We use this approximation to reproduce the shape of the distribution of $\mathcal{Q}$ for the Homodyne measurement with no Rabi drive (presented in Fig.~1 of  the main text), in Fig.~\ref{nR}.~(d).
 \end{document}